\documentclass[prd,preprint,11pt]{revtex4}

    \usepackage{amsmath,amssymb}
    \usepackage{makeidx}
    \usepackage{amsfonts}
    \usepackage[usenames,dvipsnames]{pstricks}
    \usepackage{subfigure}
    \usepackage{epsfig}
    \usepackage{slashed}
    \usepackage{enumerate}
    \usepackage{wasysym}
    \usepackage{graphicx}

\makeatletter
\newcommand{\xequal}[2][]{\ext@arrow 0055{\equalfill@}{#1}{#2}}
\def\equalfill@{\arrowfill@\Relbar\Relbar\Relbar}
\makeatother

\DeclareFontFamily{U}{mathb}{\hyphenchar\font45}
\DeclareFontShape{U}{mathb}{m}{n}{
<-6> mathb5 <6-7> mathb6 <7-8> mathb7
<8-9> mathb8 <9-10> mathb9
<10-12> mathb10 <12-> mathb12
}{}
\DeclareSymbolFont{mathb}{U}{mathb}{m}{n}
\DeclareMathSymbol{\llcurly}{\mathrel}{mathb}{"CE}
\DeclareMathSymbol{\ggcurly}{\mathrel}{mathb}{"CF}

\makeindex

\begin{document}

\preprint{IZTECH-HEP-03/2017}

\title{
Naturalizing Gravity of the Quantum Fields, and the Hierarchy Problem}

\author{Durmu{\c s} Demir\footnote{demir@physics.iztech.edu.tr}}
\affiliation{Department of Physics, {\.I}zmir Institute of
Technology, IZTECH, TR35430 {\.I}zmir, TURKEY}
\date{\today}

\begin{abstract}
It is shown that gravity can be incorporated into the Standard Model (SM) in a way solving the hierarchy problem. For this, the SM effective action in flat spacetime is adapted to curved spacetime via not only the general covariance but also the gauge invariance. For the latter,  gauge field hard masses, induced by loops at the UV scale $\Lambda$, are dispelled by construing $\Lambda$ as the constant value assigned to  curvature. This gives way to an unprecedented mechanism for incorporating gravity into the SM in that the hierarchy problem is solved by transmutation of the Higgs boson $\Lambda^2$--mass into the Higgs-curvature coupling, and the cosmological constant problem is alleviated by metamorphosis of the vacuum $\Lambda^4$--energy into the Einstein-Hilbert term. Gravity emerges correctly if the SM is accompanied by a secluded dark sector sourcing non-interacting dark matter, dark energy and dark radiation. Physics beyond the SM, containing Higgs-phobic scalars that resolve the strong CP problem, flavor problem, baryogenesis and inflation, respects the hierarchy. Majorana neutrinos are naturally incorporated if $\Lambda$ lies at the see-saw scale. This mechanism, in general, leaves no compelling reason to anticipate new particles at the LHC or higher-energy colliders. 
\end{abstract}

\maketitle

\tableofcontents

\section{Introduction and Summary}
The standard model of particle physics (SM), spectrally completed as a renormalizable
quantum field theory with the discovery of its Higgs boson at the LHC \cite{higgs-discovery}, suffers from a number
of problems related to absence of a dark matter candidate \cite{dm-review-0,dm-review},
sensitivity to ultraviolet (UV) domain \cite{natural}, difficulties in the incorporation of gravity \cite{gravity}, and various others. Traditionally, unnaturalness due to strong UV sensitivity 
has been singled out as the foremost problem to be solved.  The 
solutions (supersymmetry, extra dimensions and technicolor paradigms and their variants) generically predict new particles weighing at the Fermi scale, enjoying specific selection rules,
and interacting with the SM particles. However, the LHC
experiments, having already reached pretty above the Fermi scale, have discovered yet not a single particle, leave aside sets of particles \cite{lhc-search-exotica,lhc-search-susy}. This
aporia, unless obviated by possible discoveries at higher
luminosities, seems to summon alternative approaches. In this regard, attempts have been made to develop some novel mechanisms such as the twin Higgs \cite{twin-higgs}, cosmological relaxation \cite{relax-Higgs}, Higgs frames \cite{conformal-frame}, gravitational relaxation \cite{relax-grav}, and large copies of the SM spectrum \cite{large-copies-SM}.

The present work chooses not the UV sensitivity but the incorporation of gravity as the foremost problem to be solved. It does so because the way gravity joins the three gauge forces in the SM can facilitate the requisite mechanism needed for rehabilitating the SM in the UV. This conjecture, which was first proposed in \cite{durmus1} and \cite{durmus2} using different methods, rests on the fact that gravity, as the weakest of all forces, inherently pertains to the UV domain and can have the potential to counteract the UV sensitivity of the SM. Indeed, the SM can have its UV boundary at the gravitational scale or at an intermediate one, depending on if there exist low-energy phenomena necessitating intermediate-scale physics beyond the SM. 

Sec. \ref{Options} investigates the question of how quantized matter can gravitate. It makes the crucial observation that gravity can be incorporated into the SM by carrying the flat spacetime SM effective action into curved spacetime \cite{durmus1,durmus2}. (This is in stark contrast to the usual practice in which the SM is first placed in curved geometry and its effective action is formed thereupon.) This novel approach transforms flat spacetime effective field theories with UV scale $\Lambda$ into curved spacetime effective field theories with metric elasticity $\Lambda$. 
This mechanism, to be henceforth termed as gravitization, well applies to the SM and its known extensions in such a way that the gravitational scale $M_{Pl}$ derives from the SM UV boundary $\Lambda \lesssim M_{Pl}$ if there exists a secluded new physics (NP) sector accompanying the SM. Gravitization distinguishes between the inertial scales (like particle masses which are immune to geometry) and  gravital scales (like the UV scale $\Lambda$ which are indicative of geometry change). 

Sec. \ref{nat-prob} gives the SM effective action in flat spacetime in way explicating its naturalness problems. Those problems, the cosmological constant problem arising from the quartic UV sensitivity of the vacuum energy \cite{ccp}, the big hierarchy problem related to the quadratic UV sensitivity of the Higgs boson mass \cite{natural}, and the symmetry breakdown problem \cite{hard-break} caused by the hard UV masses of the gauge bosons, are all gravital in nature. 

Sec. \ref{gravitize-SM} provides a detailed, systematic study of how gravitization naturalizes the SM. Sec. \ref{gravitize-SM-a} proves that the symmetry breakdown problem can be solved if gravity is incorporated by replacing flat metric with a curved metric, and simultaneously, identifying the $\Lambda^2$-masses of gauge bosons with the Ricci curvature of that curved metric \cite{durmus2}. In a sense, the SM effective action is reinterpreted as the constant-curvature, flat-metric cut-view of the SM effective action in curved spacetime. This interpretation, as shown in Sec. \ref{gravitize-SM-b}, solves the UV-end of the cosmological constant problem \cite{ccp} by canalizing the quartic UV contributions to vacuum energy into not the cosmological constant but the gravitational constant (as was first realized in \cite{demir-ccp}). It solves also the big hierarchy problem \cite{natural} by transforming the quadratic UV corrections to Higgs boson mass into Higgs-curvature non-minimal coupling \cite{durmus1,durmus2}. Sec. \ref{gravitize-SM-c} shows that $M_{Pl}$ can derive from $\Lambda$ ({\it a la} Sakharov \cite{sakharov}) correctly only if there is a secluded NP sector accompanying the SM \cite{durmus1,durmus2}. Sec. \ref{gravitize-SM-d} shows that the actual UV scale of the SM must lie at the gravitational scale. Sec. \ref{gravitize-SM-e} gives a full account of the NP by emphasizing its potential to source non-interacting dark matter \cite{vilenkin-peebles} (the so-called ebony matter in \cite{durmus1,durmus2}) as well as ebony energy and ebony radiation (non-interacting limits of  dark energy \cite{v-dark-ener} and dark radiation \cite{ebony-rad}). It contrasts gravitization with supersymmetry \cite{susy-dark} and extra dimensions \cite{xtra-dim-dark} in Table \ref{tablom0} based on their dark components. Sec. \ref{gravitize-SM-e} states the eventual naturalized SM setup, and confronts it with supersymmetry \cite{susy} and extra dimensions \cite{xtra-dim} in Table \ref{tablom}. It shows that the logarithmic $M_{Pl}$ dependencies in the naturalized SM can be reinterpreted within the familiar dimensional regularization scheme \cite{cutreg-dimreg}. The setup makes it clear that gravitization leaves no ground for anticipating interacting new physics beyond the SM (BSM) at the LHC or higher-energy colliders.
It makes it also clear that possible discoveries at higher-luminosity LHC can hardly cause the little hierarchy problem \cite{lhp} because they must be coupling to the SM feebly enough to escape detection at the present precision.

Sec. \ref{BSM} investigates naturalness of the known BSM setups. They are required by various phenomena not describable by the SM. They are the strong CP problem, flavor problem, baryogenesis, inflation, and of course, the neutrino Majorana masses.  All phenomena but Majorana masses are traditionally modeled with Higgs-phobic BSM setups. Sec. \ref{BSM-phobic} shows case by case that Higgs-phobicity is technically natural and radiatively stable. Sec. \ref{BSM-philic} discusses Majorana neutrinos in the usual see-saw setup, which is Higgs-philic, and shows that the Higgs sector is stabilized by lowering $\Lambda$ to the right-handed neutrino mass scale through an even larger secluded NP sector. Sec. \ref{natural-BSM} gives the naturalized SM plus BSM setup, including the requisite secluded NP sector \cite{durmus2}. 

Sec. \ref{conc} concludes the work.

\section{Effective Field Theory in Curved Geometry or\\ Curved Geometry on Effective Field Theory?}
\label{Options} The SM is the quantum field theory of the strong and
electroweak interactions. It does not involve gravity. It must
therefore be adapted to curved spacetime to incorporate gravity as
the fourth known force. This is not as straightforward as it sounds,
however. The reason is that quantized curved spacetime is
unavailable and classical curved spacetime is uneasy for quantized
fields \cite{gravity}. (Curved momentum space \cite{gravity-2} seems 
possible if locality is sacrificed \cite{relative-locality}.) Even so, classical spacetime still serves the purpose
as far as the low-energy effective action is concerned \cite{eff-ac-grav}. Then, as the
{\bf customary approach},
\begin{itemize}
\item the SM can be directly carried into curved spacetime and its effective action can be constructed therein.
\end{itemize}
This means that the SM is valid up to the gravitational scale
$M_{Pl}$. Matter loops, deformed by spacetime curvature, lead to
linear and higher curvature terms in the SM effective action.  The
main problem with this approach is that gravity is held classical
even for loop momenta as high as the gravitational scale and this,
which is physically equivocal due to quantum gravitational effects
\cite{wilczek}, obscures the curvature sector of the SM effective
action. (There exists of course no such problem in non-curvature
sectors, which remain unchanged between the flat and curved
geometries.) To remedy the curvature sector, as an {\bf alternative approach},
\begin{itemize}
\item not the SM action but the SM effective action in flat spacetime can be carried into the classical curved spacetime \cite{durmus1,durmus2}.
\end{itemize}
This approach, in contrast to the customary approach above,
possesses a number of unprecedented features:
\begin{enumerate}

\item
 Using flat spacetime effective action is precisely what is
 needed for putting quantized matter in classical curved
 spacetime, in a  consistent way. It is so because, up to
 residual quantum fluctuations at the IR, effective action is
 essentially the classical action with quantum-corrected
 couplings, and thus, it harmoniously accord with the classical
 curved spacetime \cite{gravity,seif}. Of such is the SM effective action.

\item
 Adding curvature by hand, as in the customary approach above,
 is hampered by incalculability of the added terms since
 construction of the SM effective action in flat spacetime
 leaves behind no matter loops to induce any new interactions.
 This implies that curvature sector must be formed by a
 different mechanism that involves only the parameters of the
 flat spacetime SM effective action \cite{durmus2}.

\item
 Incorporating gravity necessitates generation of $M_{Pl}$ along
 with the formation of curvature sector. But, the SM, meant to
 hold at the Fermi scale, does, at the tree level, not possess
 any large scale of $M_{Pl}$ size. Indeed, it can possess a
 large scale as such only at the loop level if loop momenta are
 cut off at some high UV scale $\Lambda$ \cite{sakharov}. This scale, which must
 generate $M_{Pl}$ proportionally upon the incorporation of
 gravity, is necessarily a {\bf gravital scale} indicative of
 geometry change from flat to curved. (Gravital scale is what 
 characterizes the curved phase space in \cite{relative-locality}). It can, of course, not be
 an {\bf inertial scale} as masses of particles are insensitive
 to spacetime geometry. Gravitational nature of $\Lambda$
 ensures that the SM gets completed in the UV by classical
 gravity \cite{durmus1,durmus2}.

\item
 Extending the SM by new particles is necessary. The reason is that it is not possible to keep $\Lambda \leq
 M_{Pl}$ without extra degrees of freedom. Indeed, $M_{Pl}$ is
 expected to derive from $\Lambda$ as 
\begin{eqnarray}
\label{induce}
M_{Pl}\;\; = \!\!\!\!\!\sum\limits_{\begin{array}{c} {\text{SM particles,}}\\ {\text{new particles}}
\end{array}}\!\!\!\!\!\!\!\!\!\!\!\!\left({\text{loop factors}}\right) \times \Lambda
\end{eqnarray}
where the loop factors, thanks to those of the new fields,
add up close to unity to enable $\Lambda \leq M_{Pl}$ \cite{durmus1,durmus2}. Physics
details of the new particles are ascertained by the way gravity is incorporated. 

\item
 Presuming that the SM holds good up to the gravital scale
 $\Lambda$ does automatically imply that the new fields in (\ref{induce}) 
 do not interact with the SM fields. They form therefore a  secluded {\bf new physics} sector (NP), which is completely decoupled from the SM \cite{durmus2}.

\end{enumerate}
This whole mechanism can be interpreted to realize {\bf
gravitization} of flat spacetime effective field theories (in the
spirit of magnetization of a piece of iron). Indeed, all it does is to
gravitize flat spacetime effective quantum field theories having UV
scale $\Lambda$ to make them curved spacetime effective field theories
having metric elasticity proportional to $\Lambda$. Below, 
gravity will be incorporated into the SM via gravitization
(not the customary approach discussed at the top).

\section{The SM Effective Action and Its Naturalness Problems}
\label{nat-prob} 
The SM fields can be split as $\psi_{SM} +
\delta\psi_{SM}$ where $\psi_{SM}$ is the slow component having only
sub-Fermi frequencies and $\delta\psi_{SM}$ is the fast component
having trans-Fermi frequencies as high as $\Lambda$. Integrating out
$\delta\psi_{SM}$, the slow fields develop the effective action 
\begin{eqnarray}
\label{action-SM-flat} S_{\Lambda}\left(\eta\right) =  S\left({\eta}, \psi_{SM}, G^{-\!1}_{F}\log\left(G_F \Lambda^2\right)\right) + S^{0}_{\Lambda}\left({\eta}\right) +  S^{1}_{\Lambda}\left(\eta\right)
\end{eqnarray}
in the flat spacetime of metric ${{\eta_{\mu\nu}}}$ such that
$S\left({\eta}, \psi_{SM}, G^{-\!1}_{F}\log\left(G_F
\Lambda^2\right)\right)$ piece contains the tree-level $SM$ action  plus
logarithmic UV corrections plus higher-dimensional
${\mathcal{O}}\left(1/G_F\Lambda^2\right)$ terms,
\begin{eqnarray}
\label{action-UV-0} S^{0}_{\Lambda}\left({\eta}\right) = \int d^4x \sqrt{ \left\Vert\eta\right\Vert} \left\{a \Lambda^4 + b \Lambda^2 H^{\dagger} H \right\}
\end{eqnarray}
adds a UV-sized vacuum energy along with a UV-sized Higgs boson
mass, and
\begin{eqnarray}
\label{action-UV-1}
S^{1}_{\Lambda}\left(\eta\right) =\int d^4x\sqrt{ \left\Vert\eta\right\Vert} c_V \Lambda^2 \eta_{\mu\nu} {\mbox{Tr}}\!\left\{ V^{\mu} V^{\nu}\right\}
\end{eqnarray}
introduces UV-sized gauge boson masses so that hypercharge, isospin
and color are all explicitly broken at the UV.

In the main, $S_{\Lambda}\left(\eta\right)$ is an effective theory
whose all parts, including $S^{0}_{\Lambda}\left({\eta}\right)$ and
$S^{1}_{\Lambda}\left(\eta\right)$, are finite and physical. There
are no UV divergences necessitating regularization. There are
instead UV sensitivities varying from sector to sector. Quark and
lepton masses, which vary with $\Lambda$ only logarithmically, stay
close to their tree-level values. In view of the Higgs mechanism
that generate them, therefore, $S^{0}_{\Lambda}\left({\eta}\right)$
and ${{S^{1}_{\Lambda}\left(\eta\right)}}$ stand as subversive
outliers compared to $ S\left({\eta}, \psi_{SM},
G^{-\!1}_{F}\log\left(G_F \Lambda^2\right)\right)$, and give cause
to the following problems:
\begin{enumerate}[ {\text{Naturalness Problem}} 1.]
\item
 In  $S^{0}_{\Lambda}\left({\eta}\right)$, the UV-sized shift
 $\delta V = a \Lambda^4$ in vacuum energy leads to  the {\bf
 cosmological constant problem} \cite{ccp} if gravity is incorporated via
 the customary approach of Sec. \ref{Options}.

\item
 In  $S^{0}_{\Lambda}\left({\eta}\right)$ again, the UV-sized
 shift $\delta m_h^2 = - 2 b \Lambda^2$ in Higgs boson mass
 gives rise to the {\bf big hierarchy problem} \cite{natural} by lifting the SM
 towards the UV scale $\Lambda$ which, according to the
 gravitization approach, pertains to $M_{Pl}$.

\item
 In  $S^{1}_{\Lambda}\left({\eta}\right)$, the UV-sized shift
 $\delta M_V^2 = c_V \Lambda^2$ in gauge boson masses entails
 the {\bf symmetry breakdown problem} \cite{hard-break} at the UV scale.
\end{enumerate}
These problems are, in nature, {\bf gravital naturalness problems}
in that they arise from the UV scale $\Lambda$ that generates
$M_{Pl}$ as in (\ref{induce}).

\section{Gravitization of the SM Effective Action and Its Naturalization}
\label{gravitize-SM}
This section gives a systematic and in-depth analysis of the
SM effective action in regard to its naturalization via gravitization. The end product will be the naturalized SM effective action in curved spacetime in conjunction with a 
secluded NP sector. 
\subsection{Restoring Gauge Invariance}
\label{gravitize-SM-a}
In view of the Problem 3, incorporation of gravity into the SM
effective action (\ref{action-SM-flat}) proceeds in common practice
by making the change
\begin{eqnarray}
S^{1}_{\Lambda}\left(\eta\right) \xrightarrow[{\text{comma}}\ \rightarrow\ {\text{semicolon}}]{\eta_{\mu\nu}\rightarrow g_{\mu\nu}} S^{1}_{\Lambda}\left(g\right)
\end{eqnarray}
where $g_{\mu\nu}$ is a putative metric. It is by adding an explicit
curvature sector
\begin{eqnarray}
\label{change1}
S^{1}_{\Lambda}\left(g\right) \xrightarrow{{\text{add curvature}}} S^{1}_{\Lambda}\left(g\right) + \int d^4x\sqrt{ \left\Vert g \right\Vert} \left\{ M^2 R(g) + V + c_R R^2 + \dots\right\}
\end{eqnarray}
that $g_{\mu\nu}$ is ensured to be a curved metric physically
different than $\eta_{\mu\nu}$. The problem with this new action, as
already noted in Sec. \ref{Options}, is that the added parameters
$M$, $V$, $c_R$, $\dots$ are all incalculable constants. They are
incalculable simply because high-frequency quantum fluctuations that
can induce them have already been integrated out in forming the flat
spacetime effective action $S^{1}_{\Lambda}\left(\eta\right)$. This
means that curvature sector must be formed not by hand but by a
mechanism based on solely the parameters in
$S^{0}_{\Lambda}\left({\eta}\right)$ and
${S^{1}_{\Lambda}\left(\eta\right)}$. This is precisely what
gravitization is doing \cite{durmus1,durmus2}.

The goal here is to carry ${{S^{1}_{\Lambda}\left(\eta\right)}}$
into curved spacetime in a way evading the problems encountered in
(\ref{change1}). To that end, it proves exemplary to recall first
the Stueckelberg mechanism \cite{stueckelberg}
\begin{eqnarray}
\label{stueckelberg}
\int d^{4} x \sqrt{ \left\Vert\eta\right\Vert} c_V \Lambda^2 {\mbox{Tr}}\!\left\{ \eta_{\mu\nu}  V^{\mu} V^{\nu}\right\}
\xrightarrow{0 \prec S} \int d^{4}x \sqrt{ \left\Vert\eta\right\Vert}
c_V \Lambda^2 {\mbox{Tr}}\left\{\eta_{\mu\nu} \left(V^{\mu}-D^{\mu}S\right) \left(V^{\nu}-D^{\nu}S\right)\right\} && \nonumber\\
\xrightarrow{\Lambda e^{i S} \prec \Phi}
\int d^{4}x \sqrt{ \left\Vert\eta\right\Vert}  c_V  {\mbox{Tr}}\left\{\eta_{\mu\nu} \left(D^{\mu}\Phi\right)^{\dagger} D^{\nu}\Phi\right\}&&
\end{eqnarray}
which is seen to restore the gauge invariance by promoting
${{S^{1}_{\Lambda}\left(\eta\right)}}$ to a scalar field kinetic
term. In other words,  ${{S^{1}_{\Lambda}\left(\eta\right)}}$ arises
from a scalar field kinetic term cut-viewed at constant scalar field
configuration. But,  scalar fields, if not classical, cannot provide
a natural UV completion. Thence, in need of a working alternative,
it proves effectual to start with the trivial identity \cite{durmus2}
\begin{eqnarray}
{{S^{1}_{\Lambda}\left(\eta\right)}} &={\textstyle{S^{1}_{\Lambda}\left(\eta\right)}} - \int d^{4}x \sqrt{ \left\Vert \eta \right\Vert}
\frac{c_V}{2} {\mbox{Tr}}\left\{ \eta_{\mu\alpha} \eta_{\nu\beta}V^{\mu\nu} V^{\alpha\beta}\right\} +
\int d^{4}x \sqrt{ \left\Vert \eta\right\Vert}  \frac{c_V}{2} {\mbox{Tr}}\left\{\eta_{\mu\alpha} \eta_{\nu\beta}V^{\mu\nu} V^{\alpha\beta}\right\} \label{1st}
\end{eqnarray}
wherein the same kinetic structure is added to and subtracted from
${{S^{1}_{\Lambda}\left(\eta\right)}}$. This identity facilitates a
regularization scheme in which
${{S^{1}_{\Lambda}\left(\eta\right)}}$ is metamorphosed into a
gauge-invariant quantity in curved spacetime by way of the following
steps \cite{durmus2}:
\begin{eqnarray}
 {{S^{1}_{\Lambda}\left(\eta\right)}} &\xequal{\text{by-parts}}& -
 \int d^{4}x \sqrt{ \left\Vert \eta\right\Vert} \frac{c_V}{2} {\mbox{Tr}}\!\left\{\eta_{\mu\alpha} \eta_{\nu\beta}V^{\mu\nu} V^{\alpha\beta}\right\}\nonumber\\
&+& \int d^{4}x  \sqrt{ \left\Vert \eta\right\Vert} {c_V} {\mbox{Tr}}\!\left\{V^{\mu}\left(-{{D}}^2 \eta_{\mu\nu} + D_{\mu}D_{\nu} + V_{\mu\nu} + \Lambda^2 \eta_{\mu\nu} \right) V^{\nu}\right\}
\nonumber\\&+&  \int d^{4}x  \sqrt{ \left\Vert \eta \right\Vert} {c_V}  {\mbox{Tr}}\!\left\{{{D}}_{\mu} \left( \eta_{\alpha\beta} V^{\alpha} V^{\beta\mu}\right)\right\}\label{2nd}\\
&\xrightarrow[{\text{comma}} \rightarrow {\text{semicolon}}]{\eta_{\mu\nu} \prec g_{\mu\nu}}& - \int d^{4}x  \sqrt{ \left\Vert g\right\Vert} \frac{c_V}{2} {\mbox{Tr}}\!\left\{g_{\mu\alpha} g_{\nu\beta}V^{\mu\nu} V^{\alpha\beta}\right\}\nonumber\\
&+& \int d^{4}x  \sqrt{ \left\Vert g\right\Vert} {c_V} {\mbox{Tr}}\!\left\{ V^{\mu}\left(-{\mathcal{D}}^2 g_{\mu\nu} + {\mathcal{D}}_{\mu} {\mathcal{D}_{\nu}} + V_{\mu\nu} + \Lambda^2 g_{\mu\nu}\right) V^{\nu}\right\}
\nonumber\\&+&  \int d^{4}x  \sqrt{ \left\Vert g\right\Vert} {c_V}  {\mbox{Tr}}\!\left\{{\mathcal{D}}_{\mu} \left( g_{\alpha\beta} V^{\alpha} V^{\beta\mu}\right)\right\}\label{3rd}\\
&\xrightarrow[{\text{gauge invariance}}]{\Lambda^2 g_{\mu\nu} \prec R_{\mu\nu}\left({}^g\Gamma\right) }& - \!\!\int d^{4}x  \sqrt{ \left\Vert g\right\Vert} \frac{c_V}{2} {\mbox{Tr}}\!\left\{g_{\mu\alpha} g_{\nu\beta}V^{\mu\nu} V^{\alpha\beta}\right\}\nonumber\\
&+& \int d^{4}x  \sqrt{ \left\Vert g\right\Vert} {c_V} {\mbox{Tr}}\!\left\{V^{\mu}\left(-{\mathcal{D}}^2 g_{\mu\nu} + {\mathcal{D}}_{\mu} {\mathcal{D}_{\nu}} + V_{\mu\nu} + R_{\mu\nu}\left({}^g\Gamma\right)\right) V^{\nu}\right\}
\nonumber\\&+& \int d^{4}x  \sqrt{ \left\Vert g\right\Vert} {c_V}  {\mbox{Tr}}\!\left\{{\mathcal{D}}_{\mu} \left( g_{\alpha\beta} V^{\alpha} V^{\beta\mu}\right)\right\}\label{4th}\\
&\xequal{{\text{by-parts back}}}& - \int d^{4}\!x  \sqrt{ \left\Vert g \right\Vert} \frac{c_V}{2} {\mbox{Tr}}\!\left\{g_{\mu\alpha} g_{\nu\beta}V^{\mu\nu} V^{\alpha\beta}\right\}\nonumber\\&+& \int d^{4}\!x  \sqrt{ \left\Vert g \right\Vert} \frac{c_V}{2} {\mbox{Tr}}\!\left\{g_{\mu\alpha} g_{\nu\beta}V^{\mu\nu} V^{\alpha\beta}\right\}\label{5th}\\
&=& 0 \label{6th}
\end{eqnarray}
such that switching from $\eta_{\mu\nu}$ to $g_{\mu\nu}$ takes
$D_{\mu} = \partial_{\mu} + V_{\mu}$ into ${\mathcal{D}}_{\mu}=
{}^{{}^{g}\Gamma}\nabla_{\mu} + V_{\mu}$ with
${{}^{g}}\Gamma^{\lambda}_{\mu\nu}$ being the Levi-Civita
connection. This derivation withholds surface terms to make no
specific assumptions about asymptotics and involves two critical
stages:  
\begin{enumerate}
\item Equation (\ref{3rd}) at which the flat metric $\eta_{\mu\nu}$ is taken to a putatively-curved metric $g_{\mu\nu}$.  This is the common practice. 

\item Equation (\ref{4th}) at which the gauge boson mass-squared is taken to the Ricci curvature of $g_{\mu\nu}$. This is not  common practice. This unprecedented assignment
 ensures curvedness of $g_{\mu\nu}$ not by putting curvature by
 hand as in (\ref{change1}) but by reinterpreting the UV scale
 in terms of the spacetime curvature.
\end{enumerate}
These two mappings, together with the remaining steps that led to
(\ref{6th}), constitute what was called gravitization in Sec.
\ref{Options}. They lead to the following peculiarities \cite{durmus2}:
\begin{enumerate}[(a)]
\item
 Notorious ${{S^{1}_{\Lambda}\left(\eta\right)}}$, which breaks
 gauge invariance at the UV, gets carried into curved spacetime
 to disappear there and then. This solves the Naturalness
 Problem 3 mentioned in Sec. \ref{nat-prob}.

\item
 Curved geometry emerges for a symmetry reason. It is the
 restoration of gauge invariance that necessitates  curved
 spacetime.

\item
 Reading backwards from  (\ref{6th}) to (\ref{1st}), it is seen
 that  ${{S^{1}_{\Lambda}\left(\eta\right)}}$ can  be
 constructed as constant-curvature ($R_{\mu\nu}(g) \models
 \Lambda^2 g_{\mu\nu}$),  flat-metric ($g_{\mu\nu} \models
 \eta_{\mu\nu}$) cut-views of the null quantity $0 = -  \int d^{4}x \sqrt{
 \left\Vert g \right\Vert} \frac{c_V}{2} {\mbox{Tr}}\left\{
 V^{\mu\nu} V_{\mu\nu}\right\}$ $+  \int d^{4}x \sqrt{
 \left\Vert g\right\Vert}  \frac{c_V}{2}
 {\mbox{Tr}}\left\{V^{\mu\nu} V_{\mu\nu}\right\}$ in curved
 spacetime.

\item The cut-view interpretation implies that $\Lambda$ exists explicitly
also in the  curved spacetime field theory. This means that not
all occurrences of $\Lambda$ in  $S_{\Lambda}\left(\eta\right)$
are meant to be replaced by the spacetime curvature.
\end{enumerate}
These features will be decisive in naturalizing the SM via
incorporation of gravity.

\subsection{Naturalizing the Higgs and Vacuum Sectors}
\label{gravitize-SM-b}
Having done with
${\textstyle{S^{1}_{\Lambda}\left(\eta\right)}}$, it is now time to
scrutinize $S^{0}_{\Lambda}\left({\eta}\right)$ in view of its
transcendental quartic and quadratic UV sensitivities. Its curved
spacetime image ${{S}}^{0}_{\Lambda}\!\left(g\right)$, to be
constructed through the mappings (\ref{3rd}) and (\ref{4th}), must,
on physical grounds, involve {\bf no extra couplings} not found in
$S^{0}_{\Lambda}\left({\eta}\right)$ as no quantum loops are left to
induce any new couplings, {\bf no curvature-free term} as curved
geometry must disappear as soon as quantum corrections are removed,
namely, $\left.{{S}}^{0}_{\Lambda}\!\left(g,
R\right)\right\rfloor_{R = 0} =
S^{0}_{\Lambda}\left({\eta}\right)\!\Big\rfloor_{\Lambda = 0}$, and
{\bf no new forces} other than gravity as spacetime can attain
required elasticity if $\Lambda$ nears the gravitational scale
\cite{sakharov}. These physical constraints give
${{S}}^{0}_{\Lambda}\left(g, R\right)$  the familiar
Einstein-Hilbert shape \cite{durmus2}
\begin{eqnarray} \label{action-UV-0-curved} {{S}}^{0}_{\Lambda}\left(g, R\right) &=& \int
    d^4x \sqrt{ \left\Vert g \right\Vert} \left\{ \frac{a}{4} \Lambda^2 {R}(g) + \frac{b}{4}  {R}(g) H^{\dagger} H \right\}
\end{eqnarray}
such that
\begin{enumerate}
\item
 $\frac{b}{4}$ forms the non-minimal Higgs curvature coupling as
 a complete metamorphosis of the quadratic UV contributions to
 the Higgs boson mass-squared. This solves the big hierarchy
 problem (Naturalness Problem 2)  in Sec. \ref{nat-prob}.

\item
 $\frac{a}{2} \Lambda^2$ sets the fundamental scale of gravity
 as an unprecedented transmutation of the quartic UV
 contributions because they are canalized into not the
 cosmological constant but the gravitational constant. This
 saves the cosmological constant from violent UV effects. The
 cosmological constant problem (Naturalness Problem 1 in Sec.
 \ref{nat-prob}) is thus resolved only in the {\bf UV end}.

The remnant vacuum energy after gravitization gets contributions
from the Higgs rest energy by an amount $(m_H^2)^2\!
\log\left(G_F \Lambda^2\right)$, from the electroweak breaking
by an amount $\left(m_H^{2}\right)^2$, from the tree-level
action by an incalculable amount $V_0$, and from the
quark-hadron phase transition by an amount $\Lambda_{QCD}^4$.
These physically distinct contributions, bearing at most
logarithmic UV sensitivity, measure  tantalizingly bigger than the
observational value $m_{\nu}^4$. This is the {\bf IR end} of the
cosmological constant problem. It is yet to be understood.

\item $R(g)^2$, $R_{\mu\nu}(g) R^{\mu\nu}(g)$ and $R_{\alpha\mu\beta\nu}(g) R^{\alpha\mu\beta\nu}(g)$, even when put in ghost-free form, cannot appear in (\ref{action-UV-0-curved}) simply because they come with new coefficients not found in $S^{0}_{\Lambda}\left({\eta}\right)$. This property makes a case for the Einstein gravity. This unprecedented feature clears away all possible higher-curvature terms that are normally impossible to eliminate by any symmetry argument.
\end{enumerate}
The action (\ref{action-UV-0-curved}) is a concrete realization of
the gravitization approach proposed in Sec. \ref{Options}. In this
form, the SM is free from all three naturalness problems listed in
Sec. \ref{nat-prob}, and encodes all four known forces in nature.

\subsection{Generating the Gravitational Constant}
\label{gravitize-SM-c}
The main problem with  (\ref{action-UV-0-curved}) is
that, already at one loop,
\begin{eqnarray}
 \label{a-par}
 a = \frac{1}{64 \pi^2} (n_b-n_f)
 \end{eqnarray}
is negative for $n_b = 28$ bosonic and $n_f = 90$ fermionic degrees
of freedom in the SM. This means that the SM spectrum alone is not
sufficient for inducing gravity correctly. It is thus 
necessary to introduce new fields as was anticipated in (\ref{induce}). The
extra fields, which constitute a secluded NP sector,  must enable
generation of the gravitational constant \cite{durmus1,durmus2}
\begin{eqnarray}
\label{MPl}
\frac{1}{2} (a+a^{NP})\Lambda^2 = M_{Pl}^2
\end{eqnarray}
through the loop factor $a^{NP}$ that gathers contributions of all
the NP fields as in (\ref{induce}). 

\subsection{Specifying the UV Scale of the SM}
\label{gravitize-SM-d}
In general, the UV scale $\Lambda$ must lie at or below $M_{Pl}$ as no field theory can survive in the trans-Planckian domain. Its actual 
value is determined by the UV end of the underlying field theory. 
The SM, standard model of the known forces and matter with no new 
fields weighing beyond the Fermi scale, must hold good
up to the UV scale $\Lambda=\Lambda_{SM}$, where
\begin{eqnarray}
\label{SM-UV}
\Lambda_{SM} = M_{Pl}
\end{eqnarray}
as is incontrovertibly clear. This UV scale is specific to the SM. If the SM is extended by heavy fields (pertaining to neutrino Majorana masses, for instance) then their masses set a new UV scale that can be much lower than $\Lambda_{SM}$. Such high-scale extensions of the SM will be analyzed in the next section. 

\subsection{Reifying the Secluded NP Enabling the SM UV Scale}
\label{gravitize-SM-e}
The SM UV scale in (\ref{SM-UV}) leads to 
$a^{NP}\simeq 2.1$ through (\ref{MPl}). This $a^{NP}$ value translates, at one loop, into the condition \cite{durmus2}
\begin{eqnarray}
\label{sart-I}
\left(n_b^{NP}-n_f^{NP}\right)_{\Lambda_{SM}} \simeq 128 \pi^2 + 62 \approx 1325
\end{eqnarray}
implying a rather {\bf crowded} NP sector -- a fact that was
already anticipated in Sec. \ref{Options} while arguing for
gravitization. The salient features below reify the NP: 
\begin{enumerate}

\item
 The NP  {\bf does not need} to interact with the SM. This is in complete agreement with the assumption in Sec. \ref{nat-prob} that the NP {\bf
 does  not interact} with the SM. Indeed, all it has to do is to
 provide the excess bosonic degrees of freedom needed to generate the fundamental scale of gravity as in (\ref{induce}).

\item
 The spectral bound in (\ref{sart-I}) is the {\bf only
 constraint} on the NP. It is, otherwise, completely free. It
 is a generic quantum field theory with no 
 constraints on its gauge structure, particle spectrum, symmetries
 and the UV sensitivity. It can, therefore, be modeled variously like
\begin{eqnarray}
{\mathcal{G}}_{NP} = SO(51),\, SU(3)^{83},\, SU(5)^{26},\,SU(26),\, E(8)^{3},\, \dots
\end{eqnarray}
gauge theory whose particle spectrum involves gauge fields of ${\mathcal{G}}_{NP}$ plus possible scalars plus possible fermions plus possible singlets. The NP can admit numerous different modelings as there exist no restrictions other than (\ref{sart-I}).

\item
 There is one way to seclude the NP from the SM: The NP spectrum
must be composed only of the SM-singlet non-Abelian {\bf gauge
fields} $X^{a}_{\mu}$ (of a non-Abelian gauge group ${\mathcal{G}}_{NP}$) and
SM-singlet  {\bf fermions} $\chi$ (neutral or charged under ${\mathcal{G}}_{NP}$). 
This must be so because, at the renormalizable level, scalars and Abelian
gauge fields in the NP can, respectively, directly couple to the
Higgs invariant $H^{\dagger}H$ and the hypercharge field
strength tensor. The NP can therefore be described by the action
\begin{eqnarray} \label{action-flat-NP} S_{NP}\left({\eta},
\psi_{NP}\right)\! = \!\!\int\! d^{4} x \sqrt{
\left\Vert\eta\right\Vert} \left\{\!-\sum_i\frac{1}{2 g_{X_i}^2} 
{\mbox{Tr}}\left\{\eta_{\mu\alpha}\eta_{\nu\beta} X_i^{\mu\nu}
X_i^{\alpha\beta} \right\}\!  +\! \sum_j {\overline{\chi}_j} \left(i{\slashed{D}}
- m_{\chi_j}\right)\chi_j\! \right\}
\end{eqnarray}
where $i$ counts the gauge fields $X^{a}_{i\,\mu}$ of different group factors ${\mathcal{G}}_i \subset {\mathcal{G}}$, and $j$ runs over different fermion fields $\chi_i$. The secluded nature of the NP ensures that $X^{a}_{\mu}$ forms dark radiation
and $\chi$ the dark matter sector. These are pitch-dark {\bf ebony fields} completely secluded from the SM. It is, of course, possible that ${\mathcal{G}}_{NP}$ confines to form glueballs of $X^{a}_{\mu}$ and hadrons of $\chi$. The NP then becomes essentially a secluded techicolor theory with a general gauge structure. 

\item The secluded nature of the NP  is what
 distinguishes gravitization from  supersymmetry (whose NP is
 superpartners \cite{susy})  and extra dimensions (whose NP is Kaluza-Klein
 (KK) levels \cite{xtra-dim}). These three completions are contrasted in Table \ref{tablom0} in regard to
 certain NP features. One decisive feature is   the NP-SM
 coupling. This coupling, while an absolute must in extra
 dimensions and supersymmetry, does simply not exist in
 gravitization. This is a crucial point. The reason is that sfermions $\tilde{f}$
 in supersymmetry, for instance, couple
 to the Higgs field as $\lambda_{\tilde{f}H}
 {\tilde{f}}^{\dagger} {\tilde{f}} H^{\dagger} H$ with  $\lambda_{\tilde{f}H} \simeq 1$, and their
 quantum fluctuations give cause to the Higgs boson mass shift \cite{susy}
\begin{eqnarray}
\label{deltamh-SUSY}
\left(\delta m_h^2\right)_{SUSY} \propto \lambda_{\tilde{f}H} m_{\tilde{f}}^2 \log \frac{m_{\tilde{f}}^2}{\Lambda_{SM}^2}
\end{eqnarray}
which remains at acceptable level only if $G_F m^2_{\tilde{f}} \sim 1$, according to which sfermions must have already showed up at the LHC experiments. The have not, however. Pushing supersymmetry upwards with larger and larger $G_F m^2_{\tilde{f}}$ results in more and more violent destabiliztion of the Higgs sector. This problem, the {\bf little hierarchy problem} \cite{lhp}, can therefore {\bf obliterate} supersymmetry (and extra dimensions \cite{xtra-dim} and any other completion) if the NP searches at the LHC end up negative. Needless to say, as shown in Table \ref{tablom0}, gravitization is {\bf immune} to this extinctive problem.

\item
The NP can reveal itself only gravitationally. It is thus a
natural home for Dark Matter (DM). Its DM candidate is, however,
a pitch-dark {\bf Ebony Matter} (EM) having only gravitational
interactions with the SM \cite{vilenkin-peebles} (essentially invisible DM \cite{dm-review}). The EM is essentially what is indicated
by the available evidences for the DM (galactic flat rotation
curves, galaxy clusters, structure formation and gravitational
lensing). Moreover, it is fully consistent with the negative results from
various DM searches and the LHC experiments \cite{lhc-search-exotica,latest-DM-bounds}. Its constituents 
can be elementary fields as well as bound states 
\begin{eqnarray}
\label{EM}
{\text{EM}} \sim \left\{\chi_{j},\ {\text{glueballs of}}\ X^{a}_{i^{\prime}\,\mu},\ {\text{mesons or baryons of}}\ \chi_{j^{\prime}} \right\}
\end{eqnarray}
where primed (unprimed) indices refer to bound (elementary) fields. Table \ref{tablom0} lists down the DM candidates in gravitization, supersymmetry and extra dimensions (see \cite{deli} for a recent review).

\item The secluded NP sources only dark energy (DE). It forms, however, a pitch-dark {\bf Ebony Energy} (EE) having only gravitational interactions with the SM. The gauge fields $X^{a}_{i\,\mu}$ can give cause to a vector EE (essentially invisible vector DE \cite{v-dark-ener}). This is not the only possibility. Indeed, the EE can receive contributions also from the vacuum energy and possible mesons. Its constituents
\begin{eqnarray}
\label{EE}
{\text{EE}} \sim \left\{{\text{vacuum energy}},\ X^{a}_{i\,\mu},\ {\text{glueballs of}}\ X^{a}_{i^{\prime}\,\mu},\ {\text{mesons of}}\ \chi_{j^{\prime}} \right\}
\end{eqnarray}
ensure that, in gravitization, the EE arises as part of the naturalization mechanism not as a field-theoretic construct (like quintessence, k-essence and various other models) extending the vacuum energy \cite{de-models}. The gauge fields  $X^{a}_{i\,\mu}$ can also source {\bf Ebony Radiation} (essentially invisible dark radiation \cite{ebony-rad}). Table \ref{tablom0} contrasts gravitization with others in regard to their optimal DE sources. 

\end{enumerate}

\begin{table}
\caption{\label{tablom0} The NP sectors required for naturalization
of the SM in gravitization, global supersymmetry \cite{susy-dark} and extra
dimensions \cite{xtra-dim-dark}.} \footnotesize
\begin{tabular}{|l|l|l|l|l|l|}
\hline
{}&\begin{tabular}{l}Field\\ content\end{tabular}&\begin{tabular}{l}Coupling\\ to the SM?\end{tabular}&
\begin{tabular}{l}Causing the Little\\ Hierarchy Problem?\end{tabular}&\begin{tabular}{l}DE\\ candidate\end{tabular}&\begin{tabular}{l}DM\\ candidate\end{tabular}\\
\hline
\begin{tabular}{l}NP in\\ Extra Dimensions\end{tabular}&\begin{tabular}{l}KK levels of\\ graviton and\\ bulk fields\end{tabular}&Yes&Yes&
\begin{tabular}{l}Vacuum\\ Energy\end{tabular}&
\begin{tabular}{l}Lightest\\ KK Particle\end{tabular}
\\
\hline
\begin{tabular}{l}NP in\\ Supersymmetry\end{tabular}&\begin{tabular}{l}squarks,\\ sleptons\\ gauginos,\\ higgsinos\end{tabular}&Yes&Yes&
\begin{tabular}{l}Vacuum\\ Energy\end{tabular}&
\begin{tabular}{l}Lightest\\ Superpartner\end{tabular}
\\
\hline
\begin{tabular}{l}NP in\\ Gravitization\end{tabular}&\begin{tabular}{l}SM-singlet\\ non-Abelian\\ gauge fields,\\ SM-singlet\\ fermions\end{tabular}&No&No&\begin{tabular}{l}Ebony Energy\\ (in Eq.(\ref{EE}))\end{tabular}&\begin{tabular}{l}Ebony Matter\\ (in Eq.(\ref{EM}))\end{tabular}\\
\hline
\end{tabular}\\
\end{table}
\normalsize These salient features reify the NP as an integral
component of gravitization. It has the flexibility to 
source both dark energy and dark matter. Moreover, it has the
ability to naturalize the SM against both the big (by incorporating
gravity) and little (by having a secluded NP) hierarchy problems.
\begin{table}
\caption{\label{tablom} Comparison of gravitization with extra
dimensions \cite{xtra-dim} and global supersymmetry \cite{susy} in regard to the naturalization of the SM.} \footnotesize
\begin{tabular}{|l|l|l|l|}
\hline
{}&\begin{tabular}{l}Extra Dimensions\\ ($\Lambda_{SM} = {\rm TeV}$)\end{tabular}&
\begin{tabular}{l}Supersymmetry\\ ($\Lambda_{SM} \gg {\rm TeV}$)\end{tabular}&\begin{tabular}{l}Gravitization\\ ($\Lambda_{SM} = M_{Pl}$)\end{tabular}\\
\hline $\#$ of Bosons -- $\#$ of Fermions&Not Fixed&0&$\geq 128
\pi^2$\\\hline Scale of NP&TeV&TeV&Not Fixed\\\hline Nature of
$\Lambda_{SM}$&Gravital&Inertial&Gravital\\\hline Gravity&Included&Not
Included&Included\\\hline Spacetime Dimensions&$>4$&4&4\\\hline Big
Hierarchy Problem &Solved&Solved&Solved\\\hline Little Hierarchy
Problem &Not Solved&Not Solved&Irrelevant\\\hline
Tension with Negative LHC Results &Yes&Yes&No\\
\hline
Tension with Negative WIMP Searches &Yes&Yes&No\\
\hline
\end{tabular}\\
\end{table}
\normalsize

\subsection{The Naturalized SM}
\label{gravitize-SM-f}
So far, only the NP and gravity sectors have been particularized. What have been left out are the logarithmic
parts $S\left({\eta}, \psi_{SM},  G^{-\!1}_{F}\log\left(G_F
\Lambda_{SM}^2\right)\right)$  in  (\ref{action-SM-flat}) and
$S_{NP}\left({\eta}, \psi_{NP},  G^{-\!1}_{F}\log\left(G_F
\Lambda_{SM}^2\right)\right)$ following from (\ref{action-flat-NP}). These
are, however, already natural in $\Lambda_{SM}$ and can therefore be carried into
curved spacetime simply by replacing the flat metric $\eta_{\mu\nu}$
with the curved metric $g_{\mu\nu}$. There arises then the complete
action
\begin{eqnarray}
\label{setup} S^{SM+NP}\left(g,R\right) &=&  S\left(g, \psi_{SM}, G^{-\!1}_{F}\log\left(G_F M_{Pl}^2\right)\right)\nonumber\\ &+&
S_{NP}\left(g, \psi_{NP}, G^{-\!1}_{F}\log\left(G_{F} M_{Pl}^2\right)\right)\nonumber\\
&+&\!\! \int
    d^4x \sqrt{ \left\Vert g \right\Vert} \left\{ \frac{1}{2}M_{Pl}^2 R(g) -\zeta_H H^{\dagger}H R(g) \right\}
\end{eqnarray}
which forms a natural setup involving the SM plus gravity plus a secluded NP sector.  The Higgs-curvature coupling, set by the loop factor $\zeta_H \sim 10^{-2}$, is too weak to have any significance (the Einstein and Jordan frames are essentially identical \cite{faraoni}).    

This naturalized SM setup attains the usual structure after transforming it into dimensional regularization. Indeed, the
logarithmic UV dependencies in the actions $S\left({g}, \psi_{SM},
G_F^{-1}\log\left(G_{F} M_{Pl}^{2}\right)\right)$ and
$S_{NP}\left({g}, \psi_{NP}, G_F^{-1}\log\left(G_{F}
M_{Pl}^{2}\right)\right)$ can be reinterpreted as loop integrals in a
$4-\epsilon$ dimensional momentum space of total volume $\mu^{2
\epsilon} \infty^{4- 2\epsilon}$. Then, the formal equivalence \cite{cutreg-dimreg}
\begin{eqnarray}
\log\!\left( G_{F}M_{Pl}^{2}\right) = {2}/{\epsilon} + \log G_F \mu^2
\end{eqnarray}
transforms logarithmic actions into the
dimensional regularization scheme. Needless to say,
$\epsilon=1/\log\left({M_{Pl}}/{\mu}\right)$ is finite but small.
Subtractions of $1/\epsilon$ terms (as in MS or
${\overline{\text{MS}}}$ renormalizations) wipe out all occurrences
of $\log\left(G_{F} M_{Pl}^{2}\right)$ to express both SM and NP in terms of the matching scale
$\mu$. Independence of Green functions from $\mu$ leads to
the renormalization group equations.

Gravitization leaves no naturalness reasons for new particles at the Fermi scale. 
Indeed, the SM is able to get naturalized with no need to interact with any new particle. 
There is thus no naturalness reasons to expect any new particles at the LHC.  There seems
to be no phenomenological reason either. This is supported by the negative WIMP searches. In spite of all these, however, there is no guarantee that the LHC will not come 
up with a new particle. In that case, the naturalized SM setup in (\ref{setup}) continues to hold except for the inclusion of the LHC fields $\psi_{LHC}$ and except also for the Higgs mass corrections 
\begin{eqnarray}
\label{deltamh-LHC}
\left(\delta m_h^2\right)_{LHC} \propto \lambda_{LHC}\, m_{LHC}^2 \log \frac{m^2_{LHC}}{\Lambda_{SM}^2}
\end{eqnarray}
generated by the loops of $\psi_{LHC}$ having mass $m_{LHC}$ and coupling $\lambda_{LHC}$ to the Higgs field. These corrections are expected to be mild if not negligible because $\lambda_{LHC}$ must be small for $\psi_{LHC}$ to have escaped detection so far \cite{lhc-search-exotica,lhc-search-susy}. In general, for $\lambda_{LHC} \lesssim 1/\left(G_F m_{LHC}^2\right)$ the shift in (\ref{deltamh-LHC}) is unlikely to cause the little hierarchy problem. 

The naturalized SM is examined in Table \ref{tablom} in a way contrasting gravitization,
supersymmetry and extra dimensions. The table, together with Table \ref{tablom0}, makes it clear that gravitization incorporates gravity into the SM, implements loop  corrections (in dimensional regularization), predicts a secluded sector that can host dark energy and dark matter, agrees with all the existing experimental bounds, and most importantly, solves the big and little hierarchy problems.  

\section{Physics Beyond the SM and Its Naturalization}
\label{BSM}
So far, all the focus has been put on the SM. This is because all the experiments, which have long been searching energies around the Fermi
scale, have come to confirm the SM. There are, however, certain phenomena which 
fall outside the experimental reach and which necessitate new 
physics beyond the SM (BSM). Indeed,  chronic challenges like the strong CP problem, flavor problem, neutrino Majorana masses, inflation and baryogenesis require BSM at ultra high scales.  Mechanisms behind them, by construction, fall into two classes: Those that do not cause the little hierarchy problem ({\bf Higgs-phobic BSM}) and those that do cause ({\bf Higgs-philic BSM}). These two are detailed below. 

\subsection{Higgs-Phobic BSM}
\label{BSM-phobic}
Majority of the aforementioned phenomena, actually all but the neutrino Majorana masses, are modeled by utilizing heavy SM-singlet scalars. The models are meant to work at high scales in a way without destabilizing the SM Higgs sector. They are, in a sense, separated from the SM. This separation makes sense provided that it can be attained naturally. In this respect, it is worth noting that basic interactions
\begin{eqnarray}
\label{BSM-0}
S_{BSM}\left(g,\Phi_{BSM}\right)\! &\ni&\!\!\int d^4x \sqrt{ \left\Vert g\right\Vert} \left\{\!  \lambda_{H \Phi} H^{\dagger}H \Phi_{BSM}^{\dagger}\Phi_{BSM}  - m_{BSM}^2\! \left(\Phi_{BSM}^{\dagger}\Phi_{BSM}\right)^2\right\}
\end{eqnarray}
commonly exist in all BSM phenomena modelable with scalars $\Phi_{BSM}$. The problem with this setup, as already revealed via (\ref{deltamh-SUSY}) in supersymmetry, is that the Higgs boson jumps to the BSM scale due to the mass shift 
\begin{eqnarray}
\label{deltamh-0}
\left(\delta m_h^2\right)_{BSM} \propto  \lambda_{H\Phi_{BSM}} m_{BSM}^2  \log \!\frac{ m^2_{BSM} }{\Lambda^2_{BSM}}
\end{eqnarray}
generated by $\Phi_{BSM}$ loops. This large correction, unless suppressed by some natural means, can wash out the scale seperation between the SM and the BSM. They key quantity here is $\lambda_{H\Phi_{BSM}}$. This coupling, which exists in any multi-scalar field theory, must be sufficiently small
\begin{eqnarray}
\label{bound-0}
\lambda_{H\Phi_{BSM}} \lesssim \frac{1}{G_F m^2_{BSM}}
\end{eqnarray}
for the SM and BSM to stay separated. This is what {\bf Higgs-phobicity} is. It is 
a technically  {\bf natural} (having small $\lambda_{H\Phi_{BSM}}$ involves no fine-tuning) and radiatively  {\bf stable} (quantum corrections to $\lambda_{H\Phi_{BSM}}$ are proportional to $\lambda_{H\Phi_{BSM}}$ itself) criterion. 
This means that $\lambda_{H\Phi_{BSM}}$, once set as in (\ref{bound-0}), stays put at that size. In conclusion, the Higgs-phobic models do not cause the little hierarchy problem thanks to (\ref{bound-0}) and hold good, therefore, way up to the SM UV scale $\Lambda_{BSM} =\Lambda_{SM}$. Their naturalness properties are discussed below.

\subsubsection{Strong CP Problem}
Strong CP problem, unnatural smallness the QCD vacuum angle $\theta$ imposed by the bounds on the neutron electron dipole moment, enjoys a Peccei-Quinn solution \cite{pq}. Its KSVZ realization \cite{KSVZ}  
\begin{eqnarray}
\label{KSVZ-0}
S_{BSM}\! &\ni&\!\!\int d^4x \sqrt{ \left\Vert g\right\Vert} \Big\{\!  \left(\lambda_{HS} H^{\dagger}H + m_S^2\right)S^{\dagger}S + \frac{\alpha_s}{8\pi}\theta G^{c}_{\mu\nu} {\tilde{G}}^{c\mu\nu} - \left[y_{\mathcal{Q}} {\overline{{\mathcal{Q}}_L}} S {\mathcal{Q}}_R + {\text{H.C.}}\right]\Big\}
\end{eqnarray}
involves a scalar field $S$ and a heavy quark ${\mathcal{Q}}$ 
such that $H$, $S$, ${\mathcal{Q}}_L$
and ${\mathcal{Q}}_R$ possess, respectively, $0$, $+2$, $+1$ and
$-1$ units of Peccei-Quinn charge. If $S$ develops a vacuum expectation value  $\langle S\rangle =  f e^{i 2 \varphi/f}$ with $f^2\propto m_S^2$ then the action (\ref{KSVZ-0}) refines to contain a dynamical $\theta$ term
\begin{eqnarray}
\label{KSVZ-1}
S_{BSM}&\ni& \int d^4x \sqrt{ \left\Vert g\right\Vert}\left\{ \frac{\alpha_s}{8\pi} \frac{a}{f} G^{c}_{\mu\nu} {\tilde{G}}^{c\mu\nu} \right\}
\end{eqnarray}
after a Peccei-Quinn rotation to the ${\mathcal{Q}}$
quark mass basis. The Goldstone boson $\varphi$, acquiring the value
$-\theta f$ in instanton background, leads to a solution of
the {\bf strong CP problem}. The axion, defined as $a = \varphi + \theta f$, though 
couples to the SM fields feebly with a strength $1/f$,  can overclose the Universe unless $10^{9}\ {\rm GeV}\leq f \leq 10^{12}\ {\rm GeV}$ \cite{wilczek}. 
Its tiny coupling ensures that $a$ can be a viable cold {\bf DM candidate} on top of the EM provided by the secluded NP sector. 
 
The KSVZ setup makes sense only if $S$ and $H$ are sufficiently decoupled. So indeed, 
the original KSVZ proposal takes $\lambda_{HS} = 0$ \cite{KSVZ}. This decoupling is necessary for suppressing not only the loop contributions (as in (\ref{deltamh-SUSY}) and (\ref{deltamh-0}) above) but also  the tree-level contributions ($\langle S \rangle \neq 0$) to the Higgs boson mass. Thus, in the sense of (\ref{bound-0}), the tiny coupling  
\begin{eqnarray}
\label{bound-KSVZ}
\lambda_{HS} \lesssim \frac{1}{G_F m_S^2}
\end{eqnarray}
ensures that the KSVZ setup solves the strong CP problem, gives a visible DM candidate in addition to the EM provided by the secluded NP sector, naturally avoids the little hierarchy problem, and obviously, holds good up to the SM UV scale $\Lambda_{SM} \gg f$. 

\subsubsection{Flavor Problem}
Flavor problem, search for the mechanism behind the masses and mixings of quarks and leptons, arrives at a resolution via the breakdown of flavor symmetries. Its Froggat-Nielsen \cite{fn} realization
\begin{eqnarray}
\label{flavor-0}
S_{BSM}\!\! &\ni&\!\!\!\int d^4x \sqrt{ \left\Vert g\right\Vert} \Bigg\{  \left(\lambda_{H\digamma} H^{\dagger}H + m^2_{\digamma}\right)\digamma^{\dagger} \digamma-  \Bigg[\!\left(\frac{\digamma}{\Lambda_{\digamma}}\right)^{n^\ell_{ij}}\! {\overline{F}_i} H {f}_{R j}\!+\! {\text{H.C.}}\!\Bigg]\!\Bigg\}
\end{eqnarray}
involves a scalar field $\digamma$ coupling to quarks and leptons ($F$ for isospin doublets
and $f_R$ for isospin singlets). Quark and lepton Yukawa textures are generated by ${n^u_{ij}}, {n^d_{ij}}$ and ${n^\ell_{ij}}$ once the flavon $\digamma$ develops a vacuum expectation value $\langle \digamma\rangle \propto m_{\digamma}$. Then, similar to what happened in the KSVZ setup, spontaneous flavor breaking plus flavon loop corrections destabilize the SM Higgs sector unless  
\begin{eqnarray}
\label{bound-FG}
\lambda_{H\digamma} \lesssim \frac{1}{G_F m^2_{\digamma}}
\end{eqnarray}
which is known from (\ref{bound-KSVZ}) to be a natural and stable criterion. It is with this bound that the flavon dynamics works to generate quark and lepton flavor structures without disrupting the electroweak breaking.   It is thus clear that the flavor problem can be resolved by the Froggat-Nielsen setup without giving cause to the little hierarchy problem.  Needless to say, the whole setup holds good up to the SM UV scale $\Lambda_{\digamma} =\Lambda_{SM}$, with no restrictions on presumably high flavon scale  $\langle{\digamma}\rangle$.

\subsubsection{Baryogenesis}
Baryogenesis, mechanism behind the matter--antimatter asymmetry, necessitates breakdown of baryon number symmetry at high energies. Its Affleck-Dine \cite{ad} realization
\begin{eqnarray}
\label{baryo-0}
S_{BSM}\! &\ni&\!\!\int d^4x \sqrt{ \left\Vert g\right\Vert} \Big\{\!  \left(\lambda_{HB} H^{\dagger}H + m_B^2\right) B^{\dagger} B  + \lambda^{\prime}_{B} B^4\Big\}
\end{eqnarray}
involves a complex scalar $B$ such that baryon number ($B \rightarrow e^{i b} B$) and CP ($B \leftrightarrow B^{\star}$) symmetries are explicitly broken by the complex $\lambda^{\prime}_{B}$ quartic. Baryogenesis starts with a baryonic condensate $\langle  B\rangle \propto m_B$ in the early Universe.  It is necessary for depositing excess baryon number. But it can destabilize the SM Higgs sector at the tree and the loop levels unless 
\begin{eqnarray}
\label{bound-AD}
\lambda_{HB} \lesssim \frac{1}{G_F m_B^2}
\end{eqnarray}
which is a natural criterion as discussed below (\ref{bound-0}). The Affleck-Dine setup can, therefore, generate the observed matter-antimatter asymmetry without causing 
the little hierarchy problem. Obviously, the SM extended with the baryonic scalar $B$ holds good up to $\Lambda_{SM}$. The actual value of $\langle B\rangle$ depends on if the baryogenesis started during or at the end of inflation. 

\subsubsection{Inflation}
Inflation, exponential expansion of the early Universe leading to flatness and homogeneity, rests on negative-pressure sources like the vacuum energy and low-kinetic scalar fields. The Planck data \cite{obs-inf} are known to prefer models  
\begin{eqnarray}
\label{inf-0}
S_{BSM}\! &\ni&\!\!\int d^4x \sqrt{ \left\Vert g\right\Vert} \Big\{\!  \lambda_{H\Phi_I} H^{\dagger}H \Phi_I^{2} -V\left({\Phi_I}\right)\Big\}
\end{eqnarray}
with plateau potentials $V\left(\Phi_I\right)$ \cite{plato-inf} which does not blow up at large $\Phi_I$. The inflaton and Higgs scales are hierarchically separated. This separation gets, however, completely washed out at the loop level unless 
\begin{eqnarray}
\label{bound-IN}
\lambda_{H\Phi_I} \lesssim \frac{1}{G_F M_I^2}
\end{eqnarray}
when $\Phi_I$ has mass $M_I$. This bound, as discussed below (\ref{bound-0}), is small yet natural. This means that inflation can be realized naturally without causing the little hierarchy problem. It holds good way up to $\Lambda_{SM}$. 

\subsection{Higgs-Philic BSM}
\label{BSM-philic}
Certain phenomena, by their nature, can necessitate significant couplings to the Higgs field.  It then becomes a true challenge to model them in a natural manner. The  neutrino Majorana masses fall into this class. They are detailed below. 

\subsubsection{Majorana Neutrinos}
In naturalizing the SM, neutrinos were ascribed Dirac masses in order to have all fields
weighing below the Fermi scale. This is quite possible but Majorana masses are also possible. Experimental precision at present cannot tell which is realized in nature \cite{which}. It is, therefore,  necessary to determine if Majorana masses can also have a natural description.  Its see-saw setup \cite{majorana}
\begin{eqnarray}
\label{neut-0}
S_{BSM} &\ni&\int d^4x \sqrt{ \left\Vert g\right\Vert} \left\{ \lambda_{\nu} {\overline{L}} H N  + \frac{1}{2} M_N {\overline{N^c}} N\right\}
\end{eqnarray}
involves the right-handed neutrinos $N$ with masses $M_N$ and Yukawa couplings $\lambda_{\nu}$ to the lepton doublets $L = \left(\nu_L, e_L\right)$. The three right-handed neutrinos, as heavy ($M_N \simeq 10^{-5} M_{Pl}$)  {\bf Higgs-philic} ($|\lambda_{\nu}|\sim {\mathcal{O}}(1)$) states, lead to realistic neutrino masses
\begin{eqnarray}
m_{\nu} \simeq \lambda_{\nu} \left(G_F M_N\right)^{-1} \lambda_{\nu}^{\dagger}
\end{eqnarray}
when integrated out at the tree level. Integrating out them in the loops, on the other hand, leads to the Higgs boson mass shift
\begin{eqnarray}
\label{deltamh-neutrino}
\left(\delta m_h^2\right)_{\nu} \propto \lambda_{\nu} M_N^2 \log \frac{M_N^2}{\Lambda_{BSM}^2}
\end{eqnarray}
whose destabilization of the Higgs sector, due to Higgs-philicity, can be avoided only by putting the UV boundary at the right-handed neutrino masses. Indeed, setting
\begin{eqnarray}
\label{low-Lambda}
\Lambda_{BSM} = M_N
\end{eqnarray}
wipes out all the logarithmic corrections due to right-handed neutrino loops. Obviously, this can happen only if all three right-handed neutrinos are {\bf degenerate} in mass. There is clearly nothing wrong with this as because flavor structures can stem from $\lambda_{\nu}$. This strict degeneracy implies that leptogenesis may not be the mechanism behind the baryogenesis. 

Gravitization, as follows from (\ref{sart-I}), can yield the low UV scale in (\ref{low-Lambda}) with a crowded NP sector. Indeed, the UV scale reduces from $\Lambda_{SM}=M_{Pl}$ down to $\Lambda_{BSM} = M_N \approx 10^{-5} M_{Pl}$ if the secluded NP sector has 
\begin{eqnarray}
\label{sart-II}
\left(n_b^{NP}-n_f^{NP}\right)_{\Lambda_{BSM}} \simeq 128 \pi^2 10^{10}+ 62 \approx 10^{13}
\end{eqnarray}
more bosons than fermions. Obviously, the NP pertaining to the BSM is much more crowded than the one pertaining to the SM. Its crowdedness, however, poses no problem as it is completely decoupled from the BSM and serves as a natural home for the EM and the EE, as discussed in Sec. \ref{gravitize-SM-e}. 

There is a crucial question to be answered: Does the constraint (\ref{sart-II}) involve any fine-tuning in satisfying (\ref{low-Lambda})? The answer is no. The reason is that once their scale is set by $\Lambda_{BSM}$, neutrino masses attain their splittings through $\lambda_{\nu}$, in a way not different than the other Yukawas in the SM. In a sense, neutrino Majorana masses give an indirect probe of the boson-fermion number difference in the secluded NP sector. 

\subsection{The Naturalized SM + BSM}
\label{natural-BSM}
The SM plus BSM, accompanied by a secluded NP sector, is governed by the grand action 
\begin{eqnarray}
\label{setup-BSM} 
S^{SM+BSM+NP}\left(g,R\right)&=&
S\left(g, \psi_{SM}, G^{-\!1}_{F}\log\left(G_F \Lambda_{BSM}^2\right)\right)\nonumber\\ &+&
S_{NP}\left(g, \psi_{NP}, G^{-\!1}_{F}\log\left(G_{F} \Lambda_{BSM}^2\right)\right)\nonumber\\ 
&+& \int d^4x \sqrt{ \left\Vert g \right\Vert} \left\{ \frac{1}{2}M_{Pl}^2 R(g)-\zeta_H H^{\dagger}H R(g)\right\}\nonumber\\  
&+& \sum\limits_{\psi_{BSM}=S, \digamma, B, \Phi_I} S_{BSM}\Bigg(g, \psi_{SM}, \psi_{BSM}, N, m^2_{\psi_{BSM}}\log\frac{\Lambda_{BSM}^2}{m_{\psi_{BSM}}^2}\Bigg)
\nonumber\\
&-& \sum\limits_{\psi_{BSM}=S, \digamma, B, \Phi_I} \int d^4x \sqrt{ \left\Vert g \right\Vert}\; \zeta_{\psi_{BSM}}{\psi^{\dagger}_{BSM}}\psi_{BSM} R(g) 
\end{eqnarray}
such that all the interactions of the BSM fields $\psi_{BSM}$, with themselves and with the SM fields $\psi_{SM}$, are put in the action $S_{BSM}$. Their contact interactions with the curvature, set by the loop factors $\zeta_{\psi_{BSM}} \sim 10^{-2}$, are too weak to be significant (the Einstein and Jordan frames are essentially identical \cite{faraoni}).  This setup is all natural provided that  the BSM scalars $\phi_{BSM}\subset\psi_{BSM}$ couple to the Higgs field in a way respecting the bound (\ref{bound-0}). Needless to say, $\Lambda_{BSM} = M_N$ ($\Lambda_{BSM}=\Lambda_{SM}$)  if the neutrinos are Majorana (Dirac). 

Basically, all the statements in Sec. \ref{gravitize-SM-f} about the naturalized SM setup  (\ref{setup}), including the use of dimensional regularization, hold also for the SM+BSM setup (\ref{setup-BSM}). It is worth emphasizing that the gravitization leaves no compelling reason for hypothesizing new particles at the LHC or higher-energy colliders. The BSM physics arises only on  empirical or conceptual necessity, and its naturalness is evident in the cases discussed in Sec. \ref{BSM-phobic} and \ref{BSM-philic} above. 

\section{Conclusion}
\label{conc}
This work has shown that flat spacetime effective field theories with destabilizing UV sensitivities can be gravitized to metamorphose into curved spacetime effective field theories with mere logarithmic UV sensitivities. The UV scale $\Lambda$ of the flat spacetime field 
theory generates the fundamental scale of gravity  as $M_{Pl}= c_{Pl} \Lambda$, with $c_{Pl}$ controlled by a secluded field-theoretic sector. This mechanism, gravitization, has been shown to lead to a natural setup composed of the Einstein gravity without higher-curvature terms, a secluded NP sector sourcing the ebony matter and dark energy, and the renormalized SM in curved spacetime stabilized against the big hierarchy problem. The little hierarchy problem is irrelevant and the cosmological constant problem is resolved in the UV end. In contrast to extra dimensions, supersymmetry and various other UV-safe completions of the SM, gravitization does not necessitate any new fields interacting withe the SM, and hence, shows good agreement with the current results from the LHC experiments and dark matter searches.  

Gravitization leaves no compelling reason for anticipating new particles at the LHC or higher energy colliders.  Nevertheless, there are phenomena, ranging from the strong CP problem to neutrino masses, indicating that the SM needs be extended at high energies. All but neutrino 
Majorana masses are modeled with heavy scalar fields whose relevant couplings to the SM Higgs field are generically taken small as a technically natural and radiatively stable constraint. These weak couplings prevent the little hierarchy problem.  The see-saw models, on the other side,  are naturalized by bringing the UV scale down to the right-handed neutrino masses through a crowded secluded field-theoretic sector.

This work can be furthered in various aspects. First, it is necessary to solve the residual IR part of the cosmological constant problem. The solution, if any, can again be a gravitational one unless some form of new physics pops up near the neutrino mass scale. In this regard, an extension of the gravitization towards the IR could prove effectual. Second, the secluded sector can be modeled in various ways, including the methods of grand unified theories. In such concrete models, a detailed study of the ebony matter, ebony energy and ebony radiation can lead to interesting astrophysics and cosmology. Third, it could be complementary to have a proper understanding of the suppressed couplings between the Higgs and the BSM scalars. Fourth, it is important to determine experimentally if neutrinos are Majorana or Dirac. The two lead to entirely different UV descriptions. Last but not least, gravitization needs be tested against relevant phenomenological, cosmological and astrophysical phenomena.

This work is supported in part by the T{\"U}B{\.I}TAK grant 115F212.

\end{document}